# Opening practice: supporting Reproducibility and Critical spatial data science

Chris Brunsdon[1] and Alexis Comber[2]

[1] National Centre for Geocomputation, Maynooth University, Maynooth, Ireland

Email: Christopher.Brunsdon@mu.i.e.

[2] School of Geography, University of Leeds, Leeds, UK

Email: a.comber@leeds.ac.uk

**Abstract**

This paper reflects on a number of trends towards a more open and reproducible approach to geographic and spatial data science over recent years. In particular it considers trends towards Big Data, and the impacts this is having on *spatial* data analysis and modelling. It identifies a turn in academia towards coding as a core analytic tool, and away from proprietary software tools offering 'black boxes' where the internal workings of the analysis are not revealed. It is argued that this closed form software is problematic, and considers a number of ways in which issues identified in spatial data analysis (such as the MAUP) could be overlooked when working with closed tools, leading to problems of interpretation and possibly inappropriate actions and policies based on these. In addition, this paper and considers the role that reproducible and open spatial science may play in such an approach, taking into account the issues raised. It highlights the dangers of failing to account for the geographical properties of data, now that all data are spatial (they are collected some*where*), the problems of a desire for *n=all* observations in data science and it identifies the need for a critical approach. This is one in which openness, transparency, sharing and reproducibility provide a mantra for defensible and robust spatial data science.

*Out with it boldly: truth loves open dealing*
(Queen Katharine, in Henry VIII (Act 3, part I), William Shakespeare)

*An article about computational science in a scientific publication is not the scholarship itself, it is merely advertising of the scholarship. The actual scholarship is the complete software development environment and the complete set of instructions which generated the figures.* (Jon Claerbout[1])

## 1. Introduction

Notions of scientific openness (open data, open code and open disclosure of methodology), collective working (sharing, collaboration, peer review) and reproducibility (methodological and inferential transparency), have been identified as important considerations for *critical* data science and for critical *spatial* data science within the GIScience domain[2] (Singleton et al, 2016; Shannon and Walker, 2018; Nüst et al, 2018; Singleton and Arribas-Bel 2019). These ways of working are enabled through open source coding environments like Python and *R* with *GIS functionality (Singleton et al, 2016)* Shannon and Walker, 2018; Nüst et al, 2018) *linked to notebook* tools like Jupyter and *R Markdown respectively*. This paper reflects on the developments and activities using spatial data and spatial analysis, and the associated changes in the conceptualisation of how spatial analysis *should be* undertaken - from off the shelf software tools (black boxes) in a commercial GIS to coding in high level and open languages such as R or Python. These have emerged for a number of reasons: software costs, distrust of the 'black box' where data is processed without disclosure of the processing method, recognition of the scientific advantages of working in an open source environment where new methods are typically available several years before their availability in commercial software, as well as the wider practice of user-community-generated software extensions and improvements - *social (geo-) computation*, and *reproducibility*. The distrust of the black box reflects a widely held view that data analysis and research should, wherever reasonable, be capable of reproduction / replication by a third party, where reproducibility is defined is the exact duplication of the results using the same materials, and replicability means confirming original conclusions (Nüst et al, 2018), for example with new data (Kedron et al, 2019), both as a scientific credo and to avoid further 'climategates' (Campbell, 2010), in which a key issue was that the researchers involved would not release their data (or their code). The need for such transparency derives from the dangers of uncritical acceptance of black box spatial analyses, and the potential for erroneous results of such approaches precisely because "there is less of a requirement to think about the underlying processes that are being implemented" (Singleton et al, 2016, p1512) In essence, a reproducible research philosophy is one which allows all aspects of the *answer* generated by any given analysis to be tested. This is especially the case as our lives are increasingly lived and shaped by a vast amount of data - often termed *Big Data* (upper case intentional).

The emergence of Big (spatial) Data, and its problems of representativeness, bias and veracity (Laney 2001), many of which are hidden (Kitchin, 2014), provide a further motivation for openness and reproducibility. Increasingly as decisions are mediated or made by proxy using data science, data analytics and any of their variants ('AI', 'Machine Learning', etc), the citizens whose lives are influenced by such decisions have an arguable right to be aware of the process leading to these decisions, including the way in which information was analysed to support them. We know this is not currently the case, with many examples of unintended (but none-the-less entirely predictable) bias emerging from within the data science community (O'Neil 2016). Just taking one example - 'Intelligence-lead policing' is an area in which examples of bias in planning and approach proliferate (e.g. Bellamy et al,2019; Mason et al, 2019; Joh 2017b, 2017a). These biases derive from the technological, political, social and economic "data assemblages" within which big data are deployed (Kitchin and Lauriault, 2014) leading to what some have referred to as the weaponization of Big Data (Levy and Johns, 2016) due to what Iliadis and Russo (2016) describe as "the veil of openness and transparency and responsible data practices" (p3). As our lives are increasingly the subject of the results of unseen algorithms, part of the role of openness and transparency is to promote and ensure justice, equity and balance in decisions (Foot and Venne 2004). The 'if you have nothing to hide (you have nothing to fear)' argument' is misplaced here: the innocent have the potential be deemed to be guilty as a result of the inherent biases in data (Iliadis and Russo, 2016; Dalton and Thatcher, 2015), and of particular note here spatial data (Dalton et al,2016) and their evaluation.

---

[1] *http://sepwww.stanford.edu/sep/jon/reproducible.html*
[2] Here "spatial data science" describes the spatial analysis of spatial data.

In this paper we highlight the need for a **critical** spatial data science perspective, an important component of which is reproducibility (the duplication of results and inference). Whereas, historically the statistics community focused on issues of inference, without computers and with data recorded and analysed manually, openness and 'criticality' were more easily attained. However this goal has become harder to achieve as the volume of (spatial) data science activities increases and as the amount of data collected and the range of organisations and institutions involved increases. We emphasise a number of important developments and considerations that link an evolution of thinking generally within spatial analysis, away from black boxes, the need for reproducibility paradigms in spatial analysis (as with many other areas of scientific endeavour), the social movement that is the R community and the increased need for a critical eye. We consider the developments in thinking around spatial analysis, provide a brief review of recent software developments, with a focus on R, the open source statistical package, and identify a number of important consequences for data analysis generally, and spatial analysis in particular, arising from the Big (spatial) Data revolution. We suggest a number of 'good practice' approaches for informed, transparent and defensible open spatial data science that emphasise reproducibility as an important aspect of computation and need for the ability to handle big data to be accompanied by critical thinking. We assert that the movement towards open data, open source software and documentation can play a major role in this.

## 2. Spatial Analysis

In part this paper was prompted as a reflection on some earlier work (Rey and Anselin 2006) - the focus of an issue of *Geographical analysis* discussing and showcasing software to visualise, manipulate and analyse spatial data (Anselin et al, 2006; Bivand 2006; Levine 2006; Okabe, et al, 2006; Rey and Janikas 2006; Samet and Webber 2006; Xiao and Armstrong 2006). Given the journal featuring the collection of papers, it is perhaps unsurprising that emphasis was placed on software tools that applied spatial statistical models, rather than tools to manage spatial data (a less analytical aspect of GIS software). We can loosely describe the focus of these papers as *Spatial Analysis*. Since the decade or so that has passed since the publication of this special edition, the world of spatial analysis has changed. The increase in the use of data analytics in general has led to more activity in this area, both in terms of developing methods, practical usage, developing software tools and more people becoming involved with more demand for this kind of activity. Arguably the number of people whose lives are likely to be impacted by this activity has also increased notably, and the extent of these impacts is also wider. The impact of these developments changes not only the lives of the practitioners of spatial analysis, but also those of many other people.

These developments, and some others, have all led to changes in the nature of software used for spatial analysis in the timespan since the special edition of *Geographical Analysis* in 2006. In the introductory article of the collection, Rey and Anselin identify a number of stages in the development of spatial analysis software tools up to that point (Table 1).

*Table 1. Rey and Anselin's stages of spatial analysis software evolution.*

| Phase | Description |
| --- | --- |
| 1. | Appraisal of conceptual issues and consideration of required functionality. |
| 2. | Intensive development and implementation of tools followed by initial dissemination efforts. |
| 3. | Development being refined and informed by a broader community of users - this links to a number of then emergent themes in computational research. |

Relating to the last of these, they also note that "…researchers are engaging in more hands-on programming as part of their empirical analyses" attributing this to the fact that existing 'off-the-shelf' tools do not offer the functionality or flexibility required to address a widening range of problems and applications. They also attribute this to a (then) recent rise in the popularity of languages such as Python, TCL and Ruby (and it could reasonably be argued that R should be added to that list). They also note the rise in the use of open source tools - for example referring to plans to make GeoDa open source, which at the time of writing this article (2020) have come to fruition. In general these observations now seem prescient - a notable change in spatial analysis software since then has been a further shift towards using open source software, and high level programming languages, particularly R and Python. Innovative analytical or visual approaches are now often made available in either R packages, Python modules or possibly both - and although GUI-based packages exist, there is perhaps a shift away from one-off programs offering a single technique or family of techniques and working *only* via a GUI, lacking the ability to link it to a scripting language in some way.

If we are to consider the stages in the development of software tools following those listed in Table 1, perhaps an interesting further development is that stage 1 has been revisited, particularly in terms of ideas of *reproducibility*. This captures general view that data analysis carried out in research should, wherever reasonable, be capable of reproduction by a third party (Singleton et al, 2016; Wang 2016; Shannon and Walker, 2018; Nüst et al, 2018; Singleton and Arribas-Bel 2019; Kedron et al, 2019). In essence this requires that any reporting of data analysis should provide the computational procedure used and the data, in addition to a presentation of the outcome of the analysis. If reproducibility **is** seen as desirable, then consideration of functionality is altered, thus initiating a second iteration of the three stages. In particular, reproduction of computational procedures is more directly achieved when working with code – the code itself is a more reliable record of the computation that has occurred than attempting to memorise a sequence of button clicks and menu selections. We suggest that the former provides a significantly less error-prone means of reproduction. The principle of openness is also an important, and linked, issue (Iliadis and Russo, 2016; Singleton et al, 2016; Shannon and Walker, 2018; Nüst et al, 2018; Kedron et al, 2019) as well as greater engagement with the analysis (Singleton et al, 2016). In using open source code, all aspects of the software tools used may be subject to scrutiny - so that not only the analyst's scripts, but also packages they rely on, and even the construction of the programming language itself are on open view. The rise of arguments in favour of these principles, and their incorporation into stage 1 of Table 1 re-focus the required functionality to lean much more towards open source, code-based approaches.

This re-focussing is not something occurring solely in the spatial analysis community - these ideas emerge from a much broader community of data analysts, computational statisticians and other applied quantitative researchers. It is perhaps a reflection of stage 3, where spatial analysts interact with people working in other fields, that external ideas have influenced the development of spatial analysis – with the notion of development being *informed* by a broader community. Fortunately this broader community has already made efforts to develop new tools for working in a reproducible framework. Although not specifically designed for spatial analysis, these tools are sufficiently general to be used for that purpose. In particular, R Markdown (Baumer et al, 2014) via RStudio has been widely adopted by the R community, and Jupyter by Python users. Another development reflecting the general drive towards openness is the provision of repositories for software - and more broadly open and reproducible scientific reporting. For R, there is the CRAN website, and more generally sites such as GitHub, enable work to be carried out openly, with the ability for third parties to reproduce results or make use of partial outputs in their own work – with appropriate citation (Smith et al, 2016), of course.

In addition to the move towards openness, a number of other broader developments in data analysis have occurred - also impacting spatial analysis. A major one is the advent of 'Big Data' – an idea not just implying very large $n$ and large files, but also developments in the complexity of the data (a great deal of textual data - such as tweets from Twitter - has become available since 2006) – and real time delivery. Data sets are not always static, but are often now linked to an 'on all the time' feed. These emergent characteristics provide challenges to the mode of operation of software. For example, real time data sits uneasily with a standard 'read data in - analyse - display results' approach. The situation is more like an electricity meter that is permanently connected to a supply and monitors and analyses details of what flows through. In response to this, the idea of *Dashboards* as an alternative approach has become common (see Shannon and Walker, 2018).

However, this in turn provides challenges for the principle of reproducibility discussed earlier. Also, although large files are not the sole defining characteristic of Big Data, they are *one* of its characteristics, and that in itself provides challenges - often met by adaptation of approaches to work with cloud computing, distributed databases and other linked technology.

A final - and no less significant - development has been in the widespread use and influence of 'data analytics'. Although this is a somewhat vague term, in practice this often has a spatial component. Increases in use has meant that a wider group of people are likely to be affected by decisions based on the results of such analysis - and perhaps to wider realisation amongst some commentators that these outcomes, and the motivations for carrying out the analysis in the first place are worthy of study in their own right. Spatial analysts maybe need to consider the wider influence of their work, rather than imagining themselves to be cocooned in a kind of Faraday cage, isolated from all but the logic underpinning the act of analysis. Arising from these views, calls for a discipline of Critical Data Science have been published in recent years (Dalton et al, 2016; Dalton and Thatcher (2015). In a broader sense these too add to the underlying issues in stage 1 of Table 1.

## 3. R developments in support of Open Spatial Analysis

A notable characteristic of R is that it is open source. It promotes a culture of open code and discourages black boxes. An example of the problems that the latter can cause can be found in the GIS industry software standard, ESRI's ArcGIS. Some benchmarking of geographically weighted regression (Brunsdon, Fotheringham, and Charlton 1996) compared four open implementations of GWR in three R packages and one Python package, with two black box implementations: ESRI's implementation of GWR in ArcGIS and GWR4 (Nakaya et al,2014). All of the implementations were tested with the same input data. They all gave the same results except the ESRI / ArcGIS implementation (Li, 2018), and although ESRI provide help for the GWR tools, the actual coding is closed - the underlying code is not revealed. As a result, end users have no way of identifying the cause of this discrepancy. All of the R implementations (of GWR and indeed any other method) are open - ultimately the code used in the functions can be examined. In 2001 Jan de Leeuw started to articulate the need for reproducibility in research and commenting on software he noted:
*"commercial software is closed, which means that its properties have to be taken more or less on faith (which can sometimes be quite a leap of faith, compare Richard Fateman's famous review of Mathematica®)."* (De Leeuw et al, 2001, p2)
and if the Fateman citation is checked, he revealed a number of bugs and errors in functions that were part of the programme as shipped (Fateman 1992).

The background to R, the open source statistical package is well documented (R-Core-Team 2018), as are recent technical developments that support spatial analysis. These include new and consistent formats for spatial data and spatial objects (e.g. *sf*, (Pebesma 2018)), new ways of chaining sequences of commands and data operations (e.g. piping with *dplyr* (Wickham et al,2015)), improved mapping and visualisation capability (e.g. with *tmap* (Tennekes 2018) and with *ggplot2* (Wickham et al, 2016)) and the companion software tool RStudio, which provides an integrated environment for writing and running R code (RStudio-Team 2015). A number of resources have been published that cover these developments and introduce spatial data and spatial analysis in R to the novice (Lovelace et al, 2019), take the novice reader to advanced methods in spatial analysis (Brunsdon and Comber 2018) and that develop advanced methods spatial analysis in detail (Bivand et al,2008).

In many ways these developments are part of the continuing evaluation associated with any software and production process (if spatial analysis is considered as such): methods and tools improve, become more efficient and take advantage of technological developments (increased web-access for example). It should be noted that developments of this kind are not evolutions in thought or method - they do not represent new concepts in spatial analysis. Rather they are improvements in the environment within which spatial analysis takes place, and it is these developments that have allowed a new (spatial) data science culture to flourish, reflecting and supporting huge advances in open thinking, open working, sharing, open collaboration, and, ultimately, reproducibility and transparency in research and spatial analysis.

The RStudio environment for working in R has probably resulted in greater expansion of open practices than any other R-related development, particularity the inclusion since 2014 of R Markdown (Baumer et al,2014). This allows R users to embed code, analysis and data within a single document. This functionality directly supports the 'holy grail' of scientific publishing, which is to provide sufficient detail of data pre-processing and analysis such that others following the same methods can obtain identical results. Such objectives are commonly stated in requirements for methods sections of scientific papers across disciplines:
*"The method section should provide enough information to allow other researchers to replicate your experiment or study"*[3]
*"It must be written with enough information so that: (1) the experiment could be repeated by others to evaluate whether the results are reproducible, and (2) the audience can judge whether the results and conclusions are valid."* (Kallet 2004, p1229, Public Health)

The need to publishing code has been long called for:
*"Programs written by scientists may be small scripts to draw charts and calculate correlations, trends and significance, larger routines to process and filter data in more complex ways… What they have in common is that, after a paper's publication, they often languish in an obscure folder or are simply deleted. Although the paper may include a brief mathematical description of the processing algorithm, it is rare for science software to be published or even reliably preserved…"* (Barnes 2010, p753)
but it is only very recently that as a matter of course, scientific publications and journals have encouraged authors to include code and data as part of the submission process. Some have been doing this for nearly 10 years (e.g. the journal Biostatistics) and require code and data to be submitted to a designated *reproducibility editor* who tries to replicate the results. There are also a number of long standing activities and organisations for whom open code and software is central such as FOSS4G (Free and Open Source Software for Geospatial - *https://wiki.osgeo.org/wiki/FOSS4G_Handbook*) and OSGEO (The Open Source Geospatial Foundation - *https://www.osgeo.org*). The wider open movement is spreading and a number of groupings have been established within the academic community. One is the Opening Reproducible Research project (*https://o2r.info*) which at the time of writing is developing systems to support a shift from static scientific publications to dynamic ones in order to support the reuse and transparency aspects of open research as well as identifying and overcoming some of the barriers to reproducibility. Other, are adopting open principles. AGILE, (the Association of Geographic Information Laboratories in Europe) the leading European GIS grouping, now nominates a Reproducibility Chair for each annual conference, and the conference call for papers explicitly describes the inclusion of data with software or code as submission review criterion. The aim is that submissions deemed to be reproducible after review are to be awarded a 'reproducible badge'.

---

[3] Psychology *https://www.verywellmind.com/how-to-write-a-method-section-2795726*

The need for reproducibility was perhaps given its biggest push by the 'climategate' scandal around the leaking by hackers of thousands of emails from the Climatic Research Unit (CRU) at the University of East Anglia, UK in 2009, with allegations from climate change deniers of researchers cherry picking results. As noted by Merali (2010)
*"No such evidence emerged, but the e-mails did reveal another problem — one described by a CRU employee named 'Harry', who often wrote of his wrestling matches with wonky computer software…. As a general rule, researchers do not test or document their programs rigorously, and they rarely release their codes, making it almost impossible to reproduce and verify published results generated by scientific software, say computer scientists. At best, poorly written programs cause researchers such as Harry to waste valuable time and energy. But the coding problems can sometimes cause substantial harm, and have forced some scientists to retract papers."* (Merali 2010, pp775-776)

The ongoing developments of RStudio have greatly supported such transparency by embedding code and data with documents: from extending compilation of *sweave* (Leisch 2002) in terminal R to RStudio (with .*Rnw* file extensions), the introduction of the *knitr* package in RStudio (Xie 2017) and the R Markdown markup language (Xie, Allaire, and Grolemund 2018). The ability to create reproducible documents in RStudio is now one of RStudio's main attractions, and the shift from *sweave* to *knitr* with R Markdown was a game changer.

As well as supporting open and reproducible documents, RStudio allows users to easily publish their documents online and to share them more widely. The PRubs website[4], set up by RStudio to do just this, provides a resource for anyone to share their documents. In a similar vein the increased use of repositories such as GitHub[5] allow users to share collections of documents, data files, functions - even R packages circumventing the formal R repository processes. These can public or private and the contents can be revised by collaborators, with revision histories being stored (in a similar way to a wiki).

The final RStudio supported development is the ability to publish, share and even allow others to edit whole books, through the *bookdown* (Xie 2016) package. This is similar to R Markdown because it obviates the need to be familiar with LaTeX to generate documents, but furthermore allows the user to directly publish entire books to the *bookdown* website[6]. In some cases authors have left books in progress on this site for a period of time, allowing the user community to engage with the book, suggest materials to be included and test and run the code - an innovative integrated development process (e.g. Lovelace et al, 2019) - with the final 'book' often an online book.

---

[4] *https://rpubs.com – a repository for documents describing how to do things in R.*
[5] *https://github.com – a repository that supports code sharing and collaborative coding.*
[6] *https://bookdown.org – a website that supports* writing books and long-form articles in R Markdown that can be published using the bookdown R package

## 4. Big Spatial Data considerations

The recent increase in availability, sources and variety of data has been an important development in many areas of scientific activity and research. The so called *Big Data* revolution (Kitchin 2013; Kitchin and McArdle 2016) has taken place with terms such as *data analytics*, *data science* and *spatial data science* replacing *statistics*, *statistical analysis* and *spatial analysis*. One could argue that there has always been big data - too big for Excel, too big to fit on a hard drive, and so on. Workarounds are developed such as chunking code, server-side processing on Unix / Linux servers, and parallel processing. The need to address technical issues of handling large data can be viewed as an evolutionary process (with some inevitable dead-ends). However, perhaps this is 'big data' as opposed to 'Big Data' with intentional capital initials? Is Big Data more than ordinary data with a large number of observations and variables or are they something quite different? If so, this requires us to consider our approaches to analysis beyond simple considerations of size and the associated problems for statistical analysis. The huge variety of data, data formats and techniques adds to the complexity. The *caret* R package (Kuhn 2015) for example, contains some 238 methods for *machine learning* or model construction for inference and prediction. All of the questions arising generally from the consideration of Big Data have specific implications for Big (spatial) Data.

### 4.1 From Data to Spatial Data

We generate an increasing amount of digital data as part of our everyday lives[7]. Our individual and collective digital footprints are formed from the electronic records of the many transactions we undertake: from financial, to leisure activities as well as medical and social security records and flows of people through transportation networks. However Big Data are frequently difficult to handle using traditional data structures and analysis techniques. They may be, and frequently are, noisy, messy, unrepresentative, biased and mis-sampled (Laney 2001; Marr 2014; McNulty 2014). Handling them is difficult with traditional tools such as Excel spreadsheets and standard statistical approaches.

Increasingly our digital data are spatial - they are collected some*where* - and ubiquitously have some kind of locational information attached to them. This can directly describe the location of the transaction through latitude / longitude or projection coordinates, or indirectly through the address, postcode, geographic area or a unique identifier linked to location. Thus the data revolution has also resulted in deeper consideration of issues related to *geographical* and *spatial* properties of data as locational attributes can be used inputs to models themselves or used to link to other information through some kind of spatial overlay.

From a geographical perspective spatial data can be conceptualised as a three dimensional array of location, time and variable - see for example Berry (1964) and a later re-interpretation of this concept by Goodchild et al, (2007) - there are many now well established techniques for analysing such data (for example Clark et al, 1974). The challenge now relates not to techniques for spatial analysis but to the wider availability and divergence of spatial data, the ease with which the user is able to apply suites of different techniques, and the increasing use of geographic frameworks to analyse and integrate data.

---

[7] we recognise this is a *Global North* trend.

## 4.2 The MAUP

One key issue associated with spatial analysis within *Big Spatial Data* paradigm, is the need for critical consideration of the Ecological Fallacy (Robinson 1950) and its manifestation in spatial data, the Modifiable Areal Unit Problem or MAUP (Openshaw 1984a, 1984b). This has long been identified as an issue, but implies unaddressed problems for Big Data. In outline, the MAUP posits that statistical relationships, trends and correlations trends may exhibit very different properties when data are aggregated over different reporting units at different scales as they frequently are in analyses of Big Data. The MAUP describes the distortion of rate calculations and their impact on statistical analyses whether aimed at prediction or inference. It results in differences in outcomes when data or processes are summarised in some way over different spatial units, which are then used inputs to a model. The model outputs will vary when constructed from data aggregated over different areal units (Openshaw 1984a) because of differences in scale that drive the aggregation. Openshaw's original publication on the MAUP can be downloaded from *http://www.qmrg.org.uk/catmog/index.html*.

By way of example of the need for transparency/openness, consider two models constructed from the same input data but aggregated over different spatial units. The code below models house price against socio-economic and environmental data aggregated over 2 scales of census areas – Output Areas (OAs) and Lower Super Output Areas (LSOAs). There are 1584 OAs and 298 LSOAs in the study area, with the same overall summed counts of population, unemployed etc. The data contain attributes for unemployment (*unmplyd*) as proxy for economic wellbeing, the percentages of the population who are 16-24 years, 25-44 years, 45-64 years and over 65 (*u25*, *u45*, *u65* and *o65* respectively) as lifestage indicators, and the percentage of the area containing greenspaces. House price data was scraped from Nestoria (*https://www.nestoria.co.uk*) using their API (*https://www.programmableweb.com/api/nestoria*) on 30th May 2019. It contains a number of variables but here just house price in 1000s of pounds (£) as the quantity to be predicted, and latitude and longitude were used to geographically reference each house, so it could be linked to the census data. The data are shown in Figure 1.

*Figure 1. Housing data and different census areas scales, Output Area (OA) and Lower Super Output Area (LSOA).*

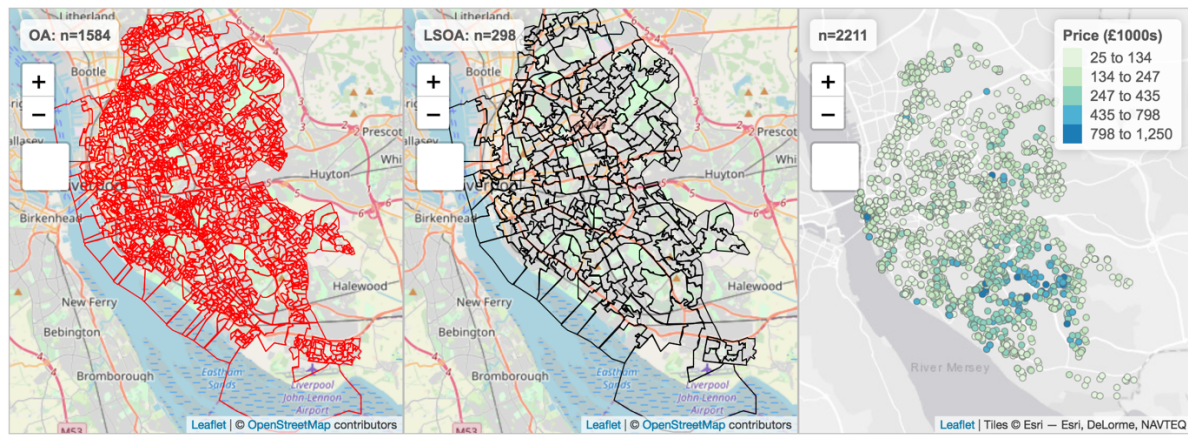

The house price data was linked to the census areas by spatial overlay so that for each property the lifestage and wellbeing variables outlined above were attached. Then simple models of house price were constructed from this data, but linked in two ways - the first using variables linked geographically at the UK Census Output Area (OA) level, the second at the larger scale Lower Super Output Area (LSOA) level. The coefficient estimates and the variable importance from the two models are shown in Tables 2 and 3.

Table 2. The model coefficient estimates for the individual input variables summarised over different scales of census area: Output Area (OA) and Lower Super Output Area (LSOA).

| Covariate | OA | LSOA |
| --- | --- | --- |
| (Intercept) | 33.986 | -43.505 |
| gs_area | 0.875 | 0.412 |
| u25 | 1.977 | 2.882 |
| u45 | 0.714 | 1.962 |
| u65 | 5.368 | 5.543 |
| o65 | 3.481 | 6.967 |
| unmplyd | -8.246 | -10.850 |

Table 3. The variable importance (expressed as a percentage), of the models constructed from data summarised over different scales of census area: Output Area (OA) and Lower Super Output Area (LSOA).

| Covariate | OA | LSOA |
| --- | --- | --- |
| (Intercept) | 25.610 | 0.000 |
| gs_area | 17.326 | 12.935 |
| u25 | 0.000 | 3.070 |
| u45 | 37.353 | 15.192 |
| u65 | 31.789 | 35.560 |
| o65 | 100.000 | 100.000 |
| unmplyd | 25.610 | 0.000 |

The results in these tables highlight the MAUP. The first issue is that the coefficient estimates for the variables are very different. For example in the OA model, each additional 1% of the *u25* variable is associated with an increase of ~£2000 in prices, whereas for the LSOA model this is associated with an increase of ~£2900. Similar statements can be made for all the covariates, as well as for the different model fits and estimates of significance. When variable importance is examined, unemployment (*unmplyd*) comes out on top for both models, but local greenspace area makes a very different contribution to explaining the variation in house price for each of the two models. The practical issue of the MAUP here is that different conclusions may be drawn depending on the spatial scale of the census variables. This is not to say that either analysis is 'wrong' - but that perhaps further consideration of the most appropriate scales to look at the census variables is needed. Another important message brings the problem of 'black boxes' once more into view - reporting regression results without stating which level of aggregation was used is only revealing part of the story. Once again this suggests that an open and reproducible approach is necessary for a meaningful interpretation of the analysis.

### 4.3 *n=all?*

As well as modelling process, the nature of the data need to be considered. One of the characteristics of Mayer-Schoenberger and Cukier's (2013) vision of big data and data sciences relates to their idea that *n=all*. If this statement was literally true, suppose an entire population's data had been collected (all the houses in Liverpool and not just the ones from the website that was scraped in May 2019), and the aim was to test a null hypothesis that the number of panels of the front door influenced house price - and that these are typically 3 kinds (4, 6 or 8 panels). Given that data for all houses and their attributes are present, one can simply split the data into 3 and compute the average price for each group. If on doing this there was a £100 difference in favour of 6 panel doors, a significance test would almost certainly reject an $H_0$ of no difference ($n$ would be huge and the standard error of the estimated difference tiny). However, is the actual difference of any consequence? In this case, the answer may well be 'no'. In part this is due to a poorly formulated hypothesis - perhaps the question here should have been 'is the 6-panel difference greater than £150?' or some questions of *practical* significance. Perhaps the *size* of the effect (here measured in UK currency) is important, rather than a very specific statement about the price difference being true or false. In this case it is context rather than a statistical test that determines what is 'significant'.

There are many more issues with the '*n*=all' idea. Meng (2016) compared analyses of a 1% survey with 60% response rate and a self-reported administrative dataset covering 80% of the population. They found that $|\rho| < 0.0034$ in order to trust the latter, indicating that Big Data even with quite slight bias is the worst option, challenging the unquestioned belief of quantity over quality and that if $n$ is very large, then hypothesis testing is infallible, and suggesting the potential for erroneous decision making and policy implementation. Here, full disclosure of the testing procedure is important, in addition to a critical view, avoiding being seduced into trusting results from a large data sample, without knowledge of the data collection process.

### 5. Critical Spatial Data Science

Critical data science perspectives are well articulated in the literature. Key to these is the need to explicitly account for and articulate the inherent assumptions and biases present in all data . This is described by Dalton et al, 2016) as '*the idea that the "we" of those who emit data is a statistically representative "we"* ' (p4). These may be due to the epistemological coverage of where we measure (the corporate "*data fumes*" are greater in the global North according to Dalton et al, (2016)) as well as the longstanding ontological choices about what and how we measure the world (Comber et al, 2004; Kedron et al, 2019), that together define what are called *the assemblages* of Big Data. A critical data science approach is one that promotes common good, *informed* data analysis and data justice (Iliadis and Russo, 2016).

However, there is a growing recognition is much of the attention that Big Data and data science have attracted over the last few years pertains to spatial data. Thus, it seems that spatial data analysts and geographers are starring in this media show whether they planned to or not. They are playing a key role in a nascent discipline that builds on earlier disciplines, and whose emergence is very much in the public gaze. However, as with data analytics and statistics, there is a danger that the emergent discipline ignores some important lessons learned from its predecessors. A number of suggestions have been made about what a critical geographical data science perspective, in the context of Big Data, CyberGIS and GIScience, would include:
- Bidirectional interactions between GIScience and Data Science to co-inform activities (Singleton and Arribas-Bel 2019);
- An ever evolving and "open" process (where 'open' is both a verb and an adjective) through which GIScience methods that are both actively transparent and actively inclusive for a range of audiences (Shannon and Walker, 2018);
- The replication of *inferences* that are robust to changes in data or spatial context (Dodge, 2019);
- "*More meaningful and spatially aware analytics (involving spatial indexing, spatial autocorrelation, spatial heterogeneity, geographic visualization, geovisual analytics and spatialization) (Singleton and Arribas-Bel 2019)*" (Dodge, 2019, p2);

If geography has become a key player here, it is vital that it ensures core critical considerations beyond these are given their due, which many geographers are aware of.

This is because spatial attributes cannot be treated as just another variable in standard statistical analyses. Many environmental, socio-economic and physical processes, such as captured in spatial data, exhibit some form of spatial pattern. For inference (in other words process understanding) one of the important assumptions is the need for independence between observations in the random parts of a model. This is frequently violated in spatial data which exhibits spatial heterogeneity and spatial autocorrelation. Consideration of the spatial aspects of data (Spatial data) and the associated need to accommodate location in statistical analyses (Spatial analysis) have been core activities for geographers for many years. A recent refinement to the 'space is (statistically) special' argument has been advanced by Comber and Wulder (2019), who, drawing from recent developments in remote sensing, argue for the importance of matching process temporal phases to the spatially and temporally aggregated phases of data, in the context of the intended analysis as well as conventional spatial considerations.

Additionally, the *ecological fallacy* (Robinson 1950) and the *MAUP* will not go away simply because one uses state of the art machine learning algorithms and spatially aggregated data, as demonstrated in the previous section. Rather, *how* such geographical and spatial consideration are addressed in analyses need to be open to scrutiny, further highlighting the need for transparency and reproducibility.

We have also argued for the need for a shift to open tools, libraries and code (and data) to enable reproduction as a fundamental principle to be taken into account in the evolution of software tools for spatial analysis in the current era of big data. Our argument is that the GIScience and spatial data science community need to pay special attention to openness, transparency, code and data sharing, and reproduction. This has a number of implications:
- Everyone does need to learn to code. It is no longer sufficient for a GI Scientists to just work with a standard GIS interface: menus, buttons and black boxes. We live in an era when a powerful GIS, data and operation instructions can be freely downloaded by someone with no spatial, geographical or GIScience background training or understanding of spatial data issues, and used to generate sophisticated analyses. Critical understandings of the inputs, methods and the results are needed alongside the analysis;
- Domain journals and conference series should require code and data to be submitted to a designated reproducibility editor who tries to replicate the results. This is a longstanding practice in many areas of science. There is no reason this cannot be done in ours;
- As an extension to the above, perhaps Geographic Data / GI Science should promote a data and analysis commentary as an adjunct to any map, visualisation or result of spatial analysis, that describes the spatial provenance of the results. This would be a sort of information quality metadata in the manner proposed by Comber et al, (2008) and Wadsworth et al, (2008) in the context of data integration;

As well as considering the *techniques* and *practice* of spatial data analysis critically, there is a need for a critical consideration of the assumptions currently surrounding (spatial) data science. Kitchin (2014), for example challenges the notion of Big Data and data science as objective and supporting all-encompassing analyses (Iliadis and Russo 2016), and instead promotes the idea of *critical data studies* (CDS) (Kitchin and Lauriault 2014). CDS encourages us to consider the common good and social contexts (Iliadis and Russo 2016) as an important consideration within the practice of Big Data and data science and extending this to take critical views on the technical aspects of the data analysis. Quoting O'Neil and Schutt (2014) *"We'd like to encourage the next-gen data scientists to become problem solvers and question askers, to think deeply about appropriate design and process, and to use data responsibly and make the world better, not worse."* A core concern relates to the wider implications of misinterpreting the inferential and predictive models *"Even if you are honestly skeptical of your model, there is always the chance that it will be used the wrong way in spite of your warnings"* (O'Neil and Schutt 2014, p354). Extending these to the geographic domain is to consider the spatial design and spatial scale of process and how the caveats associated with spatial uncertainties (Kedron et al, 2019) should be included with spatial outputs.

## 6. Concluding remarks

An increasing amount of data is being generated and therefore available for statistical analysis. Data typically have some spatial or locational attribute rendering them suitable for spatial manipulations either to summarise them over different geographies (such as census areas, administrative areas, 5km grids), or to combine them with other spatial data. Choices made about how this is done will result in very different outcomes as illustrated by the house price example above.

Further considerations related to the need for openness, transparency and reproducibility in spatial analyses arise because of the ease with which it can be undertaken. Tools and skills for spatial analysis of spatial data are no longer the preserve of geographers. Many disciplines now routinely use some kind of spatial analysis or GIS-related functionality and spatial or statistical analyses can be trivially applied, supported by the increased spatial functionality in R and RStudio and other high level programming environments, but also evident in the continued development of the QGIS project (software, training materials but little theory). However, this only increases the potential for naive and potentially incorrect spatial analyses of the kind set out here. Thus, rather than asserting that $n = $ all, if we were to provide a dictum mixing words and equations, we suggest:
Spatial Data + Spatial Analysis − Critical *Spatial* Understanding = Misinformed Decision Making

Thus there are a number of issues that should be prominent when working with spatial data and Big (spatial) Data more generally that should be part of a critical approach to data analysis and an awareness of the practical considerations of working with spatial data. These are promoted by adopting a culture of openness and reproducibility, within which any data analysis is open to review, comment, suggestion, criticisms and in many cases revision. Thus sharing code, using repositories, writing open code libraries and tools provides a set of rubrics for transparent and reproducible research. Many of these issues are generally applicable to data analytics, but some are deeply grounded in spatial data and spatial analysis, and the importance of the latter will only increase as data become *ubiquitously* spatial. Openness, transparency, sharing and reproducibility provide a mantra for defensible and robust spatial data science.

**Acknowledgements**
The data and R code used to generate the analyses in Figure 1 and Tables 2 and 3 are available from *https://github.com/lexcomber/OpeningPractice/*. This research was supported by funding from Natural Environment Research Council, Grant/Award Number: NE/S009124/1.